# STATUS OF COLDDIAG: A COLD VACUUM CHAMBER FOR DIAGNOSTICS

S. Gerstl*, T. Baumbach, S. Casalbuoni, A. W. Grau, M. Hagelstein, D. Saez de Jauregui,
Karlsruhe Institute of Technology (KIT), Karlsruhe, Germany,
C. Boffo, G. Sikler, BNG, Würzburg, Germany,
V. Baglin, CERN, Geneva, Switzerland,
M. P. Cox, J. C. Schouten Diamond, Oxfordshire, England,
R. Cimino, M. Commisso, B. Spataro, INFN/LNF, Frascati, Italy,
A. Mostacci, Rome University La Sapienza, Rome, Italy,
E. J. Wallén, MAX-lab, Lund, Sweden,
R. Weigel, Max-Planck Institute for Metal Research, Stuttgart, Germany,
J. Clarke, D. Scott, STFC/DL/ASTeC, Daresbury, Warrington, Cheshire, England,
T. W. Bradshaw, STFC/RAL, Chilton, Didcot, Oxon, England,
R. M. Jones, I. R. R. Shinton, University Manchester, Manchester, England

*Abstract*

One of the still open issues for the development of superconducting insertion devices is the understanding of the beam heat load. With the aim of measuring the beam heat load to a cold bore and the hope to gain a deeper understanding in the beam heat load mechanisms, a cold vacuum chamber for diagnostics is under construction. The following diagnostics will be implemented:
i) retarding field analyzers to measure the electron energy and flux,
ii) temperature sensors to measure the total heat load,
iii) pressure gauges,
iv) and mass spectrometers to measure the gas content.

The inner vacuum chamber will be removable in order to test different geometries and materials. This will allow the installation of the cryostat in different synchrotron light sources. COLDDIAG will be built to fit in a short straight section at ANKA. A first installation at the synchrotron light source Diamond is foreseen in June 2011. Here we describe the technical design report of this device and the planned measurements with beam.

## INTRODUCTION

Superconductive insertion devices (IDs) have higher fields for a given gap and period length compared with the state of the art technology of permanent magnet IDs. This technological solution is very interesting for synchrotron light sources since it permits to increase the brilliance and/or the photon energy at moderate costs. One of the key issues for the development of superconducting IDs is the understanding of the beam heat load to the cold vacuum chamber.

___________________
* stefan.gerstl@kit.edu

Possible beam heat load sources are:
1) synchrotron radiation,
2) resistive wall heating,
3) electron and/or ion bombardment,
4) RF effects.

The values of the beam heat load due to synchrotron radiation and resistive wall heating have been calculated and compared for the cold vacuum chambers installed at different light sources with the measured values. The disagreement between beam heat load measured and calculated is not understood [1, 2, 3].

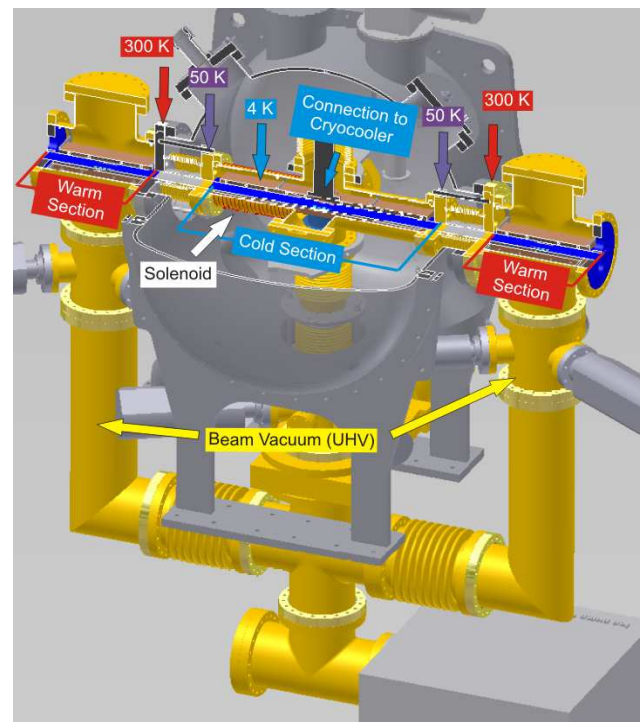

Figure 1: Sketch of the cryostat

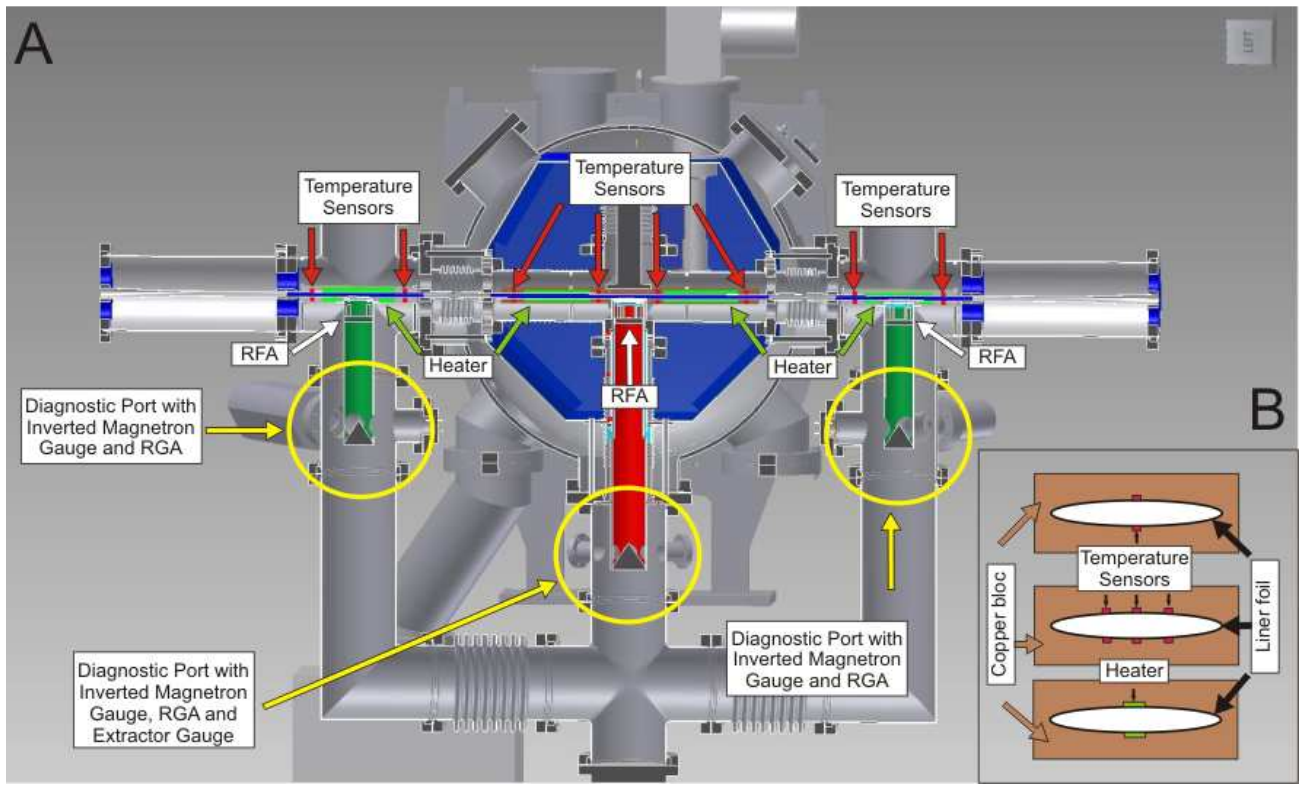

Figure 2a, b: Overview of the diagnostics installed in COLDDIAG

Studies performed with the cold bore superconducting undulator installed at the synchrotron radiation source ANKA suggest that the main contribution to the beam heat load is due to secondary electron bombardment. The electron bombardment model appears to be consistent with the beam heat load and the dynamic pressure rise observed for bunch lengths of about 10 mm [2]. Low energy electrons (few eV) are accelerated by the electric field of the beam towards the wall of the vacuum chamber, introducing heat to the cold liner and causing non-thermal desorption of gas from the cryogenic surface. In order to gain a quantitative understanding of the problem and to find effective remedies we have designed a cold vacuum chamber for diagnostics [3] together with the company Babcock Noell to be installed in synchrotron light sources. The goal is to measure the heat load, the pressure, the gas content and the flux and spectrum of the low energy electrons bombarding the wall.

A first installation in the Diamond light source is foreseen for June 2011. In the following we describe the technical design report of this device and the planned measurements with beam.

## THE VACUUM CHAMBER

COLDDIAG consists of a cold vacuum chamber located between two warm sections (fig. 1), one upstream and one downstream. This will allow to observe the influence of synchrotron radiation on the beam heat load and to make a direct comparison of the cryogenic and room temperature regions.

The electron beam will go through a liner designed to be exchangeable. The first liner that will be tested at Diamond will be a stainless steel foil 100 µm thick plated with 30 µm of copper, embedded in copper bars for thermal stabilization. For the diagnostic port in the cold section a design based on the COLDEX device [4] was chosen. The diagnostic devices are connected to the cold liner through a warm tube to avoid gas condensation along the path between liner and diagnostic device.

The vacuum system of COLDDIAG consists of two volumes: the isolation vacuum and the beam vacuum (UHV). A DN100 CF stainless steel 6-way cross separates the two vacua (fig. 1). The beam vacuum (fig.1 marked in yellow) includes all the diagnostic devices as well as the liner. This volume will be pumped with a 500 l/min ion pump. In addition during cold operation the cold bore of the chamber works like a cryopump. A base pressure of about $10^{-11}$ mbar is expected in the UHV vacuum in absence of the beam. The usage of bellows allows for the thermal shrinkage of the cold parts as they cool down. The bellows are optimized to minimize the heat transfer. They are also equipped with RF-fingers.

In order to simulate the liner of superconducting insertion devices the liner must be cooled down to reach 4.2 K in absence of beam. COLDDIAG is cryogen-free and cooled by a Sumitomo RDK-415D cryocooler. The system has 3 temperature regimes: 300 K, 50 K at the radiation shield, and 4 K at the liner. Calculations of thermal radiation and conductance show a heat load of 44.4 W on the shield at 50 K, and 0.65 W on the cold liner at 4.2 K. With a maximum cooling power of 35 W at 50 K at the 1$^{st}$ stage of the cryocooler and 1.5 W at 4.2 K at the 2nd stage one cryocooler is anticipated to be enough to obtain 4.2 K on the liner in absence of beam.

In order to suppress the low energy electrons bombarding the wall, a solenoid on the beam axis producing a maximum field of 100 Gauss will be wound around on one of the long arms of the cold UHV cross (symbolized by the red spiral in fig. 1).

## DIAGNOSTICS

The following diagnostics will be implemented:
i) temperature sensors to measure the total heat load,
ii) retarding field analyzers to measure the electron flux,
iii) pressure gauges,
iv) and mass spectrometers to measure the gas composition.

In total 38 temperature sensors will allow to monitor the status of the chamber and measure the beam heat load. 8 calibrated Hereaus C220 PT100 sensors in each of the warm sections and 16 calibrated Lakeshore Cernox 1050-SD in the cold section are placed directly on the liner foil (fig. 2a, b) by spring loaded screws. This gives us the possibility to obtain not only the beam heat load but also the heat distribution on the liner. To calibrate the temperature sensors to the beam heat load we use 10 ceramic heaters, which will be placed directly on the liner foil to simulate the heating from the beam (fig. 2a, b).

In each of the connection pipes, between the liner and the diagnostic ports a small half moon shaped retarding field analyzer (RFA) will be placed to obtain the electron flux of the electrons impinging the wall. During calibration of a similar RFA for the ANKA storage ring at INFN in Frascati it turned out, that it is not possible to obtain the electron distribution with the current setup [5]. This might be due to a background of secondary electrons produced inside the RFA. At the moment we are testing an improved setup using a lock-in technique. To solve the problem of the secondary electrons produced in the RFA we use an AC voltage inductively coupled to the retarding grid and detect the signal on the collector plate. With this setup we can directly acquire the first derivative of the electron current on the collector, which gives us the electron energy distribution.

At each of the three diagnostic ports, an inverted magnetron gauge (minimal measurable pressure $10^{-10}$ mbar) and a residual gas analyzer will be installed, to monitor the total pressure and the gas composition of the beam vacuum. The middle diagnostic port will also be equipped with an extractor gauge, which is more sensitive than the inverted magnetron gauges (minimal measurable pressure $10^{-11}$ mbar).

Through the middle diagnostic port also it will be possible to inject different gases. Therefore a heated high precision all metal leak valve, which allows to control the gas flow to the chamber down to $10^{-10}$ mbar*l/s was chosen. To study the effect of the injected gases under controlled conditions a defined amount of molecules must be homogeneously deposited on the cold liner foil. To do so we warm up the liner with the installed heaters to about 150 K. This will clean the liner surface as the boiling point of the most gases in the storage ring is below this temperature. Together with the cool down of the liner the gas will be injected until cryopumping of the cold surface starts. At this point, which will show up as a dramatic pressure decrease, the gas injection will be stopped. In several offline tests the amount of deposited gas will be checked by warming up the liner again and measuring the pressure increase of the adsorbed gas with the RGA's.

## PLANNED MEASUREMENTS

During normal user operation the temperature, electron flux, pressure and gas composition will be monitored to collect statistics. During machine physics at Diamond we plan to change:
1) the average beam current to compare the beam heat load data with synchrotron radiation and resistive wall heating predictions,
2) the bunch length to compare with resistive wall heating predictions,
3) the filling pattern in particular the bunch spacing to test the relevance of the electron cloud as heating mechanisms,
4) beam position to test the relevance of synchrotron radiation and the gap dependence of the beam heat load,
5) inject different gases naturally present in the beam vacuum ($H_2$, CO, $CO_2$, $CH_4$) to understand the influence of the cryosorbed gas layer on the beam heat load and eventually identify the gases to be reduced in the beam vacuum.

## SUMMARY

In this paper we reported about the design and the foreseen diagnostic devices of COLDDIAG. First tests of the experimental setup will take place at ANKA after delivery in summer 2010. A first installation of COLDDIAG at an electron storage ring is planned for June 2011 in the Diamond light source.

## REFERENCES

[1] E. Wallèn, G. LeBlanc, Cryogenics 44, 879 (2004).
[2] S. Casalbuoni et al., Phys. Rev. ST Accel. Beams 10, 093202 (2007).
[3] S. Casalbuoni et al., Proceedings of EPAC08.
[4] V. Baglin et al., Vacuum 73, 201-206 (2004).
[5] D. Saez de Jauregui et al., Proceedings of PAC 2009, Vencouver, Canada.